\documentclass[epj]{svjour}
\usepackage{graphics}

\begin{document}
\title{Collective Resonances in the Soliton Model Approach 
to Meson--Baryon Scattering}
\author{H. Weigel 
}                     
%
%
\institute{Fachbereich Physik, Siegen University,\,
D--57068 Siegen, Germany}
\date{Received: date / Revised version: date}
%
\abstract{
The proper description of hadronic decays of baryon resonances 
has been a long standing problem in soliton models for baryons.
In this talk I present a solution to this problem in the three flavor
Skyrme model that is consistent with large--$N_C$ consistency conditions. 
As an application I discuss hadronic pentaquark decays and show that 
predictions based on axial current matrix elements are erroneous.
} 
\maketitle
\section{Introduction}

Commonly hadronic decays of baryon resonances are described 
by a Yukawa interaction of the generic structure
\begin{equation}
{\mathcal L}_{\rm int}=g\,\bar{\psi}_{B^\prime}\,\phi\,\psi_B\,,
\label{eq:yukawa}
\end{equation}
where $B^\prime$ is the resonance that decays into baryon $B$ and meson 
$\phi$ and $g$ is a coupling constant. It is crucial that this 
interaction Lagrangian is \emph{linear} in the meson field.

The situation is quite different in soliton models that are based
on action functionals of only meson degrees of freedom, 
$\Gamma=\Gamma[\Phi]$. These action functionals contain
classical (static) soliton solutions, $\Phi_{\rm cl}$, that are 
identified as baryons. The interaction of these baryons with mesons
is described by the (small) meson fluctuations about the 
soliton: $\Phi=\Phi_{\rm cl}+\phi$. By pure definition we have
\begin{equation}
\frac{\delta \Gamma[\Phi]}{\delta \Phi}\Big|_{\Phi=\Phi_{\rm cl}}=0\,.
\label{eq:soliton}
\end{equation}
Thus there is no term linear in $\phi$ to be associated with 
the Yukawa interaction, eq.~(\ref{eq:yukawa}). This puzzle has become 
famous as the Yukawa problem in soliton models.  However, this does not 
mean that hadronic decays of resonances cannot be described in soliton 
models. Rather they have to be extracted from meson baryon scattering 
amplitudes, just as in experiment. In soliton models two--meson processes 
acquire contributions from the second order term 
\begin{equation}
\Gamma^{(2)}=\frac{1}{2}\,\phi\,
\frac{\delta^2 \Gamma[\Phi]}{\delta^2 \Phi}\Big|_{\Phi=\Phi_{\rm cl}}\,
\phi\,.
\label{eq:secorder}
\end{equation}
This expansion simultaneously represents an expansion in $N_C$, 
the number of color degrees of freedom: $\Gamma=\mathcal{O}(N_C)$, 
$\Gamma^{(2)}=\mathcal{O}(N_C^0)$ while terms $\mathcal{O}(\phi^3)$
vanish in the limit $N_C\to\infty$. This implies that $\Gamma^{(2)}$ 
contains all large--$N_C$ information about hadronic decays of 
resonances. We may reverse that statement to argue about computations 
of hadronic decay widths in soliton models: Their results and those 
obtained from $\Gamma^{(2)}$ \emph{must} be identical in the 
limit $N_C\to\infty$. Unfortunately, the most prominent baryon 
resonance, the $\Delta$, becomes degenerate with the nucleon as 
$N_C\to\infty$. It is stable in that limit and its decay is not 
subject to the above described litmus--test. The situation is 
more interesting when extending soliton models to flavor $SU(3)$.
In the so--called rigid rotator approach (RRA) that generates baryon 
states as (flavor) rotational excitations of the soliton, resonances 
emerge that dwell in the anti--decuplet representation of flavor $SU(3)$. 
The most discussed (and disputed) such state is the $\Theta^+$ pentaquark 
with zero isospin and strangeness $S=+1$. In the limit $N_C\to\infty$ 
the anti--decuplet states maintain a non--zero mass difference with 
respect to the nucleon. Therefore the decay properties of $\Theta^+$ 
as predicted in any soliton model must also be seen in the $S$--matrix 
for koan--nucleon scattering as computed from $\Gamma^{(2)}$. In the 
$S=-1$ sector the resulting equations of motion for $\phi$  yield a 
$P$-wave bound state whose occupation serves to describe the ordinary 
hyperons, $\Lambda$, $\Sigma$, $\Sigma^*$, etc.\@. Therefore this 
treatment of hyperon states is called the bound state approach (BSA). 
The above discussed litmus--test requires that the BSA and RRA give 
identical results for the $\Theta^+$ properties as $N_C\to\infty$. 
This did not seem to be true and it was argued that the prediction of 
pentaquarks would be a mere artifact of the RRA~\cite{It04}. Here we 
will show that that conclusion is premature and that pentaquark states 
do indeed emerge in both approaches. Furthermore the comparison between 
the BSA and RRA provides an unambiguous computation of pentaquark widths: 
It differs substantially from previous approaches based on assuming 
pertinent transition operators for $\Theta^+\to KN$~\cite{Di97,El04}. 
Details of these studies are contained in ref.~\cite{Wa05} and 
ref.~\cite{We96} may be consulted for a review on $SU(3)$ soliton models. 

\section{Constrained fluctuations and $\Theta^+$ width}

When restricted to modes spanned by the soliton's rigid rotations, the 
$P$--wave fluctuations in strangeness direction have two bound states with 
eigenenergies
\begin{equation}
\omega_{\pm}=\frac{1}{2}\left[\sqrt{\omega_0^2+\frac{3\Gamma}{2\Theta_K}}
\pm\omega_0\right]
\label{eq:omegapm}
\end{equation}
where $\Theta_K$ is the moment of inertia for the rotation of the soliton 
into strangeness direction and $\Gamma$ is the functional of the soliton 
that measures flavor symmetry breaking. The latter would be zero if the 
masses of the strange and non--strange quarks were equal. Both functionals 
are $\mathcal{O}(N_C)$. The subscript on $\omega_\pm$ refers to the 
strangeness quantum number of the bound state. Hence $\omega_0=N_C/(4\Theta_K)$ 
removes the degeneracy between $S=\pm1$ baryons. This contribution stems 
from the Wess--Zumino term in $\Gamma[\Phi]$. In accordance with the above 
discussion $\omega_\pm=\mathcal{O}(N_C^0)$. While $\omega_{-}$ is the energy 
of the above mentioned bound state describing ordinary hyperons, $\omega_{+}$ 
is eventually utilized to construct pentaquark states. 

\begin{figure}[t]
\centerline{
\resizebox{0.4\textwidth}{0.2\textheight}{\includegraphics{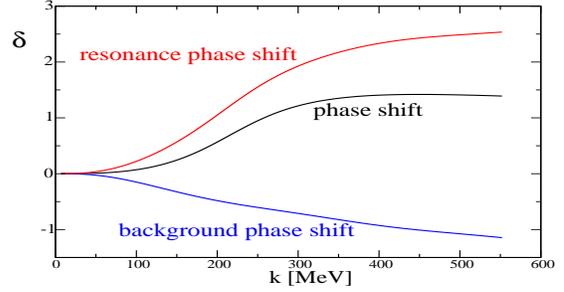}}}
\caption{\sl \footnotesize
Phase shift computed in the BSA (middle, black line) and the resonance phase 
(top, red line) shift after removal of the background (bottom, blue line)
contribution in the RVA.}
\vspace{-0.4cm}
\label{fig_1}
\end{figure}

In the RRA the collective coordinates $A(t)\in SU(3)$ that parameterize the 
flavor orientation of the soliton are canonically quantized. The resulting
Hamiltonian is (numerically) exactly diagonalized for arbitrary 
$N_C$~\cite{Wa05} and symmetry breaking~\cite{Ya88}. 
The so--computed mass difference between the states 
that for $N_C=3$ correspond to the $\Lambda$ ($\Theta^+$) and the nucleon 
approaches $\omega_{-}$ ($\omega_{+}$) as $N_C\to\infty$. This suggests
that indeed the RRA and BSA are identical in that limit. This identity
has a caveat when the restriction that BSA modes are spanned by the
rigid rotation is removed. Though $\omega_{-}<m_K$ still corresponds to a 
true bound state, $\omega_+$ is a continuum state. Thus, a pronounced 
resonance structure would be expected in the BSA phase shift around
$\omega=\omega_+$. Unfortunately, that is not the case, as seen from  
fig.~\ref{fig_1}. The BSA phase shift hardly reaches $\pi/2$ rather 
than quickly passing through this value. The ultimate comparison requires 
to generalize the RRA to the rotation--vibration approach~(RVA)
\begin{equation}
U(\vec{x\,},t)=A(t) \xi_0(\vec{x})\,
{\rm exp}\left[\frac{i}{f_\pi}\sum_{\alpha=4}^7\lambda_\alpha
\widetilde{\eta}_\alpha(\vec{x\,},t)\right]
\xi_0(\vec{x})\, A^\dagger(t)\,,
\label{Usu3}
\end{equation}
where $\xi_0(\vec{x})={\rm exp}\left[i\hat{\vec{x}}\cdot\vec{\tau}
F(|\vec{x}|)/2\right]$ is the chiral field representation of the soliton
($\Phi_{\rm cl}$) and $A(t)\in SU(3)$ parameterizes the collective 
rotations. Modes that correspond to the collective rotations must be 
excluded from the fluctuations $\widetilde{\eta}$, {\it i.e.\@} the 
fluctuations must be orthogonal to the zero--mode $z\sim {\rm sin}(F/2)$. 
Imposing the corresponding constraints for these fluctuations (and their 
conjugate momenta) yields integro--differential equations listed in 
ref.~\cite{Wa05}. In the BSA $A(t)$ is restricted to $SU(2)$ but there 
is no constraint on the fluctuations $\eta$. The above discussed 
litmus--test requires that the scattering data computed from 
$\widetilde{\eta}$ and $\eta$ to be identical when $N_C$ is sent to 
infinity.

For the moment let's omit the coupling between $\widetilde{\eta}$ and 
the collective soliton excitations. This truncation defines the background 
wave--function $\overline{\eta}$ (also orthogonal to the zero mode). 
Treating $\overline{\eta}$ as an harmonic fluctuation provides the  
background phase shift shown as the blue curve in fig.~\ref{fig_1}. 
Remarkably, the difference between the phase shifts of $\overline{\eta}$ 
and $\eta$ clearly exhibits a distinct resonance structure. This is the 
resonance phase shift to be associated with the $\Theta^+$ pentaquark in 
the limit $N_C\to\infty$!

The parameterization, eq.~(\ref{Usu3}) is not a solution to the 
classical equation of motion, The strategy rather is to solve them order 
by order in the $N_C$ expansion. Hence the arguments deduced from 
equation~(\ref{eq:soliton}) do not apply and an interaction Hamiltonian 
that is linear in the fluctuations indeed emerges. This generates 
Yukawa couplings between the collective soliton excitations and the 
fluctuations $\widetilde{\eta}$. In ref.~\cite{Wa05} we have derived 
this Hamiltonian keeping all contributions that survive as $N_C\to\infty$. 
The corresponding Yukawa exchanges extend the integro--differential 
equations for $\overline{\eta}$ by a separable potential $V_Y$, therewith 
providing the equations of motion for $\widetilde{\eta}$. For $N_C\to\infty$
the equation of motion for $\widetilde{\eta}$ is solved by 
$\widetilde{\eta}=\eta-\langle z|\eta\rangle z$. The phase shifts extracted 
from $\eta$ and $\widetilde{\eta}$ are identical because $z(|\vec{x}|)$ 
is localized in space.
Thus the BSA and RVA yield the same spectrum and are indeed equivalent 
in the large $N_C$ limit. But, the RVA provides a distinction between 
resonance and background contributions to the scattering amplitude.
Applying the $R$--matrix formalism on top
of the constrained fluctuations $\overline{\eta}$ shows that $V_Y$
\emph{exactly} contributes the resonance phase shift shown
in fig.~\ref{fig_1} when the Yukawa coupling is computed for
$N_C\to\infty$. This identifies the exchange of a state predicted
in the RRA which thus is no artifact. In contrast, pentaquarks
are also predicted by the BSA; just well hidden. Nevertheless, collective 
coordinates are mandatory to obtain finite $N_C$ corrections to the BSA
for the properties of~$\Theta^+$. Though not all ${\cal O}(1/N_C)$
operators were included in ref.~\cite{Wa05}, subleading effects are
substantial. For example, in the case $m_K=m_\pi$ the mass difference with 
respect to the nucleon increases by a factor two from $\omega_0$ to 
$(N_C+3)/4\Theta_K$ for $N_C=3$. 

\begin{figure}[t]
\centerline{
\rotatebox{270}{\resizebox{0.25\textwidth}{0.35\textheight}
{\includegraphics{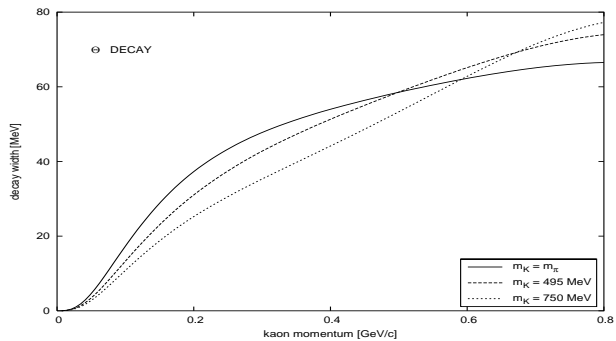}}}}
\caption{\sl\footnotesize
Skyrme model prediction for the decay width, $\Gamma(\omega)$
of $\Theta^+$ for $N_C=3$ as function of the kaon momentum
$k=\sqrt{\omega^2-m_K^2}$, {\it cf.\@} eq.~(\ref{widthsb}). 
Top: $m_K=m_\pi$, bottom: $m_K\ne m_\pi$.}
\vspace{-0.1cm}
\label{fig_2}
\end{figure}

The separable potential $V_Y$ also yields the general expression 
for the width as a function of the kaon energy
$\omega_k=\sqrt{k^2+m_K^2}$ from the $R$--matrix formalism~\cite{Wa05}
\begin{eqnarray}
\Gamma(\omega_k)&=&2k\omega_0
\left|X_\Theta\int_0^\infty r^2dr\, z(r)2\lambda(r)
\overline{\eta}_{\omega_k}(r)\right.\cr
\hspace{1cm}&&\left.
+\frac{Y_\Theta}{\omega_0}\left(m_K^2-m_\pi^2\right)
\int_0^\infty r^2dr\,z(r)\overline{\eta}_{\omega_k}(r)\right|^2\,.
\hspace{1cm}
\label{widthsb}
\end{eqnarray}
Here $\overline{\eta}_{\omega_k}(|\vec{x}|)$ is the P--wave projection 
of the background wave--function $\overline{\eta}$ for a prescribed
energy $\omega_k$ and $\lambda(|\vec{x}|)$ is a radial function that 
stems from the Wess--Zumino term. The matrix elements ($X_\Theta$ and 
$Y_\Theta$) of the collective coordinate operators that enter 
eq.~(\ref{widthsb}) are computed from the eigenstates of the 
collective coordinate Hamiltonian. The resulting width is shown for 
$N_C=3$ in fig.~\ref{fig_2} for both the flavor symmetric case and the 
physical kaon--pion mass difference.
As function of momentum, there are only minor differences between these 
two cases. Assuming the observed resonance to be the (disputed) 
$\Theta^+(1540)$ a width of roughly $40{\rm MeV}$ is read off from 
fig.~\ref{fig_2}~\cite{Wa05}. However, fig.~\ref{fig_1} suggests that 
the resonance should be about $200{\rm MeV}$ above threshold which 
corresponds to $M_{\Theta^+}\approx1.65{\rm GeV}$. Hence it seems 
very unlikely that chiral soliton models predict a light and very 
narrow pentaquark, though the numerical results for masses and widths 
of pentaquarks are model dependent.

\section{Conclusion}

In this talk I have presented a thorough comparison~\cite{Wa05} between 
the bound state (BSA) and rigid rotator approaches (RRA) to chiral 
soliton models in flavor $SU(3)$. Though I have only considered the simplest 
such model, the actual analysis merely concerns the treatment of 
kaon degrees of freedom. Therefore the qualitative results are valid 
for \emph{any} chiral soliton model.

A sensible comparison with
the BSA requires the consideration of harmonic oscillations in the
RRA as well. They are incorporated via the rotation--vibration 
approach (RVA), however constraints must be implemented to ensure that 
the introduction of such fluctuations does not double--count any 
degrees of freedom. The RVA clearly shows that the prediction of 
pentaquarks is not an artifact of the RRA, pentaquarks are genuine within 
chiral soliton models. Only within the RVA chiral soliton models generate 
interactions for hadronic decays. Technically the derivation of this 
Hamiltonian is quite involved, however, the result is as simple as 
convincing: In the limit $N_C\to\infty$, in which the BSA is undoubtedly 
correct, the RVA and BSA yield identical results for the baryon spectrum 
and the kaon--nucleon $S$-matrix. This identity also holds when flavor 
symmetry breaking is included. This demonstrates that collective coordinate 
quantization may be successfully applied regardless of whether or not
the respective modes are zero--modes. 

In the flavor symmetric case the interaction Hamiltonian contains only a 
\emph{single} structure ($X_\Theta$ in eq.~(\ref{widthsb})) of $SU(3)$ 
matrix elements for the $\Theta^+\to KN$ transition. Any additional $SU(3)$ 
structure only enters via flavor symmetry breaking. This proves earlier 
approaches~\cite{Di97,El04} incorrect that adopted any possible 
structure that would contribute in the large $N_C$ limit and fitted 
coefficients from a variety of hadronic decays under the assumption 
of $SU(3)$ relations. The study presented in this talk thus 
shows that it is not worthwhile to bother about the obvious arithmetic 
error in ref.~\cite{Di97} that was discovered earlier~\cite{We98,Ja04}
because the conceptual deficiencies in such width calculations are 
more severe. Assuming $SU(3)$ relations
among hadronic decays is not a valid procedure in chiral soliton
models. The embedding of the classical soliton breaks $SU(3)$ and 
thus yields different structures for different hadronic transitions. 

Even in case pentaquarks turn out not to be what some recent
experiments have suggested, they have definitely been very beneficial
in combining the bound state and rigid rotator approaches and
solving the Yukawa problem in the kaon sector; both
long standing puzzles in chiral soliton models.

\section{Acknowledgments}
I am very appreciative to Hans Walliser for the very fruitful 
collaboration on which this talk is based. I am grateful to the 
organizers for this pleasant conference. Attendance of the conference
has been made possible by the DFG under contract We 1254/12-1.

\end{document}